\documentclass[%
 aip,
 amsmath,amssymb,
 reprint,%
]{revtex4-1}

\usepackage{graphicx}
\usepackage{dcolumn}
\usepackage{bm}

\usepackage[utf8]{inputenc}
\usepackage[T1]{fontenc}
\usepackage{mathptmx}
\usepackage{etoolbox}
\usepackage[usenames,dvipsnames]{color}
\makeatletter
\def\@email#1#2{%
 \endgroup
 \patchcmd{\titleblock@produce}
  {\frontmatter@RRAPformat}
  {\frontmatter@RRAPformat{\produce@RRAP{*#1\href{mailto:#2}{#2}}}\frontmatter@RRAPformat}
  {}{}
}%
\makeatother
\begin{document}

\preprint{AIP/123-QED}

\title{Unsupervised identification of crystal defects from atomistic potential descriptors}
\author{Lukáš Kývala}
\affiliation{
Faculty of Physics, University of Vienna, 1090 Vienna, Austria }%
\affiliation{
Vienna Doctoral School in Physics, University of Vienna, 1090 Vienna, Austria}
\author{Pablo Montero de Hijes}%
\affiliation{
Faculty of Physics, University of Vienna, 1090 Vienna, Austria }%
\affiliation{
Faculty of Earth Sciences, Geography and Astronomy, University of Vienna, 1090 Vienna, Austria}
\author{Christoph Dellago}
\affiliation{
Faculty of Physics, University of Vienna, 1090 Vienna, Austria }%

\date{\today}

\begin{abstract}
Identifying crystal defects is vital for unraveling the origins of many physical phenomena. Traditionally used order parameters are system-dependent and can be computationally expensive to calculate for long molecular dynamics simulations. Unsupervised algorithms offer an alternative independent of the studied system and can utilize precalculated atomistic potential descriptors from molecular dynamics simulations. We compare the performance of three such algorithms (PCA, UMAP, and PaCMAP) on silicon and water systems. Initially, we evaluate the algorithms for recognizing phases, including crystal polymorphs and the melt, followed by an extension of our analysis to identify interstitials, vacancies, and interfaces. While PCA is found unsuitable for effective classification, it has been shown to be a suitable initialization for UMAP and PaCMAP. Both UMAP and PaCMAP show promising results overall, with PaCMAP proving more robust in classification, except in cases of significant class imbalance, where UMAP performs better. Notably, both algorithms successfully identify nuclei in supercooled water, demonstrating their applicability to ice nucleation in water.

\end{abstract}

\maketitle

%
%
%
%
%

Machine learning (ML) has transformed the computational modeling of atomic interactions, a cornerstone of molecular dynamics (MD) simulations. Current machine learning potentials (MLPs) allow us to reliably simulate large systems and long timescales with ab initio accuracy \cite{behler2021machine}. 
 Beyond the fitting of potential energy surfaces, the development of ML tools has also benefited the analysis of atomistic simulations, providing valuable structural information. 
For instance, supervised learning approaches based on neural networks have been used to identify different phases in polymorphic systems \cite{geiger2013neural,fulford2019deepice,dijkstra2021predictive,defever2019generalized,ishiai2023graph, ishiai2024novel,kuroshima2024machine} as well as dynamical processes \cite{huang2022machine}.
However, characterizing local atomic environments of crystal defects is a challenging task \cite{mosquera2023identifying,wang2024identification,gorfer2024structure}. Conveniently, the development of MLPs has come along with refined features describing local atomic environments, such as the Smooth Overlap of Atomic Position (SOAP) descriptors \cite{bartok2013representing} or the atom-centered symmetry functions \cite{Behler2007}, which are applied in this work.
Designing order parameters can be complex and computationally intensive, especially if no structural classification is available in advance. Therefore, unsupervised ML methods have emerged as a promising path towards discovering relevant structural features in complex systems \cite{ceriotti2019unsupervised}. 
Instead of trying to learn the local chemical environment from typically tens to hundreds of dimensions of atomic descriptors, it is common practice in unsupervised methods to initially reduce the dimensionality to two or three dimensions. Several methods for dimensionality reduction have been applied in the context of simulations \cite{frickenhaus2009efficient,teodoro2003understanding,baba2022prediction,spiwok2011metadynamics,sgourakis2011atomic,glielmo2021unsupervised}.
In particular, unsupervised classification tasks have been performed using i) topological graph order parameters combined with both Principal Component Analysis (PCA) \cite{chapman2022efficient} and diffusion maps  \cite{reinhart2017machine,reinhart2018automated}, ii) Gaussian mixture models together with both PCA \cite{spellings2018machine} and neural-network-based autoencoders \cite{boattini2019unsupervised,zhang2023fastff}, iii) a band structure encoding along with the t-distributed stochastic neighbor embedding (t-SNE) \cite{nunez2019exploring}, iv) simple three-body features \cite{reinhart2021unsupervised} jointly with the Uniform Manifold Approximation and Projection (UMAP) \cite{UMAP}, and v) SOAP combined with UMAP \cite{offei2022high,di2023zundeig,donkor2024beyond} and the Multidimensional Scaling Algorithm \cite{kruskal1964multidimensional}. \\
     
To effectively analyze MD trajectories, it is crucial to employ algorithms that meet three fundamental criteria: (1) minimal computational overhead, which can be achieved by utilizing precomputed atomistic potential descriptors; (2) distinct separation of clusters for simple classification; and (3) robustness without variations in hyperparameters and initialization. Topological graph methods fail to satisfy the requirement of minimal computational overhead due to the additional calculations involved in generating graph-based order parameters \cite{chapman2022efficient}. Furthermore, metric-preserving algorithms such as PCA and autoencoders often produce distributions with overlapping clusters, lacking clear separation, as demonstrated in this work. Graph-based algorithms like t-SNE and UMAP are known to be sensitive to variations in hyperparameters and initialization procedures \cite{Wattenberg2016,phdthesis, Becht2018}. The novel graph-based algorithm Pairwise Controlled Manifold Approximation Projection (PaCMAP) has been reported to be more robust and superior compared to t-SNE and UMAP in toy problems \cite{PaCMAP} and single-cell transcriptomic data analysis \cite{Huang2022}.

Our study compares three unsupervised dimensionality reduction techniques for identifying crystal defects: PCA, UMAP, and PaCMAP. Our goal is to validate the reported superiority of PaCMAP over the established UMAP method. We maintain consistent hyperparameters across all studies to evaluate their robustness when cluster identities are unknown beforehand. The only exception is the initialization process, which has been shown to be crucial for achieving reliable results. We select two standard benchmark systems: Si and H$_2$O. First, we assess the methods' ability to classify phases and then focus on locating point defects. For silicon, we also identify surface atoms, whereas for H$_2$O, we detect a hexagonal ice nucleus surrounded by supercooled water, an essential feature for studying ice nucleation \cite{piaggi2022homogeneous,montero2023minimum}. A detailed methodology for PaCMAP is provided, as this algorithm is less well-known than PCA and UMAP. Additionally, we discuss the nonparametric nature of PaCMAP and our approach to addressing it.


\section{Results}
\subsection{Silicon}
\subsubsection{Phase classification}
\begin{figure*}[ht]
\centering
\includegraphics[width=0.7\textwidth]{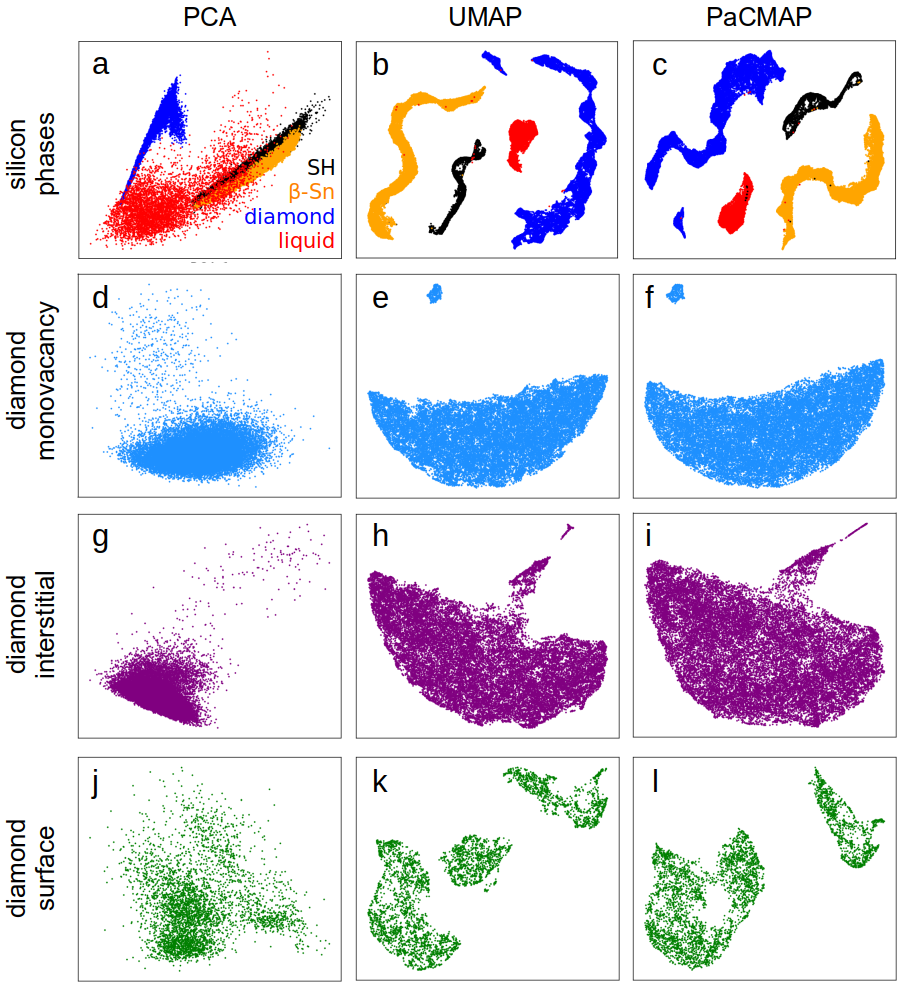}
\caption{
\textbf{Comparison of clustering results using PCA, UMAP, and PaCMAP algorithms for silicon.} Included are silicon phases (a-c), diamond monovacancy (d-f), diamond interstitial (g-i), and diamond surface (j-l).
}
\label{fig:silicon}
\end{figure*}
Silicon crystallizes in a diamond structure under standard conditions. When subjected to pressure, the diamond structure of silicon transforms into the $\beta$-Sn structure at approximately 10 GPa, the orthorhombic structure (space group Imma) at around 13 GPa, and the simple hexagonal structure at around 16 GPa \cite{Voronin2003}. High-pressure structures share the same unit cell, where atoms are positioned at (0, 0, 0), (0.5, 0.5, 0.5), (0.5, 0, 0.5 + $\nu$), and (0, 0.5, $\nu$). A $\nu$ value of 0.25 corresponds to the $\beta$–Sn structure, while the SH structure is characterized by $\nu$ = 0.5. The Imma phase provides a continuous transition between these phases. Consequently, the $\beta$–Sn and SH phases closely resemble each other, and their differentiation may not always be trivial. It is confirmed by the PCA analysis in Fig. \ref{fig:silicon}a, where these phases overlap, and a clear boundary cannot be unequivocally determined. Similarly, distinguishing the liquid phase from the solid is not achievable through PCA, as the liquid embedding overlaps with other solid phases, as already observed in water \cite{monserrat2020liquid}.

Our preliminary experiments revealed that PaCMAP and UMAP highly depend on the initialization process. Specifically, using UMAP's default spectral initialization resulted in correct clustering (Fig. \ref{fig:silicon}b) but also artificial cluster creation (Fig. S1a, in Supplementary information) and misclassification of parts of the SH phase as $\beta$-Sn (Fig. S1b). These results were obtained with identical hyperparameters, only varying the random seed. Similar results are observed with PaCMAP random initialization. To address this instability, we switched to PCA initialization. Following this change, both algorithms successfully separated the liquid phase from the solid phases and clearly distinguished between $\beta$-Sn and SH phases, as shown in Figs. \ref{fig:silicon}b and \ref{fig:silicon}c. The misclassification rates for UMAP and PaCMAP were 0.10\% and 0.08\%, respectively.

\subsubsection{Diamond point defects}
The subsequent analysis aims to localize monovacancies and interstitials within silicon's diamond structure. Monovacancies can be identified by locating the four nearest neighbors that would have formed bonds with the missing atom if it had been present. The PCA plot in Fig. \ref{fig:silicon}d shows two distributions without clear separation, whereas UMAP and PaCMAP clustering in Figs. \ref{fig:silicon}e and \ref{fig:silicon}f clearly distinguish atoms near monovacancies from those in other regions. Graph-based algorithms accurately identify the four nearest neighbors for 89\% of structures, while the remaining cases exhibit variations (3, 5, and 6 neighbors) caused by local distortions introduced by interstitials. Nonetheless, the algorithms successfully locate the monovacancy in all tested cases.

Identifying interstitial atoms is straightforward when their identity remains constant throughout the simulation. However, additional analysis is required if an interstitial atom changes its identity over time. While PCA provides no useful information (Fig. \ref{fig:silicon}g), UMAP and PaCMAP categorize silicon systems with interstitial defects into two interconnected and one separated cluster, as shown in Figs. \ref{fig:silicon}h and \ref{fig:silicon}i. The largest cluster includes atoms far from the defect, while the middle interconnected cluster comprises atoms near the interstitial. The smallest clusters are composed of atoms not arranged on a diamond lattice. Figure S2 presents examples from the smallest cluster, showing local chemical environments with varying degrees of distortion.

\subsubsection{Surface} 
We conclude our silicon analysis by examining (001) diamond surfaces using a relatively small dataset comprising 29 structures. The PCA analysis reveals only overlapping distributions (Fig. \ref{fig:silicon}j). While UMAP divides the local environments into three clusters (Fig. \ref{fig:silicon}k), PaCMAP identifies only two (Fig. \ref{fig:silicon}l). Upon examining one of the surface structures depicted in Fig. S3, we find that the UMAP clusters primarily correspond to bulk atoms, surface atoms, and first-neighbors to surface atoms. PaCMAP misses the first neighbors to surface atoms.

\subsection{Hexagonal ice}
\subsubsection{Phase classification}
\begin{figure*}[ht]
\centering
\includegraphics[width=0.7\textwidth]{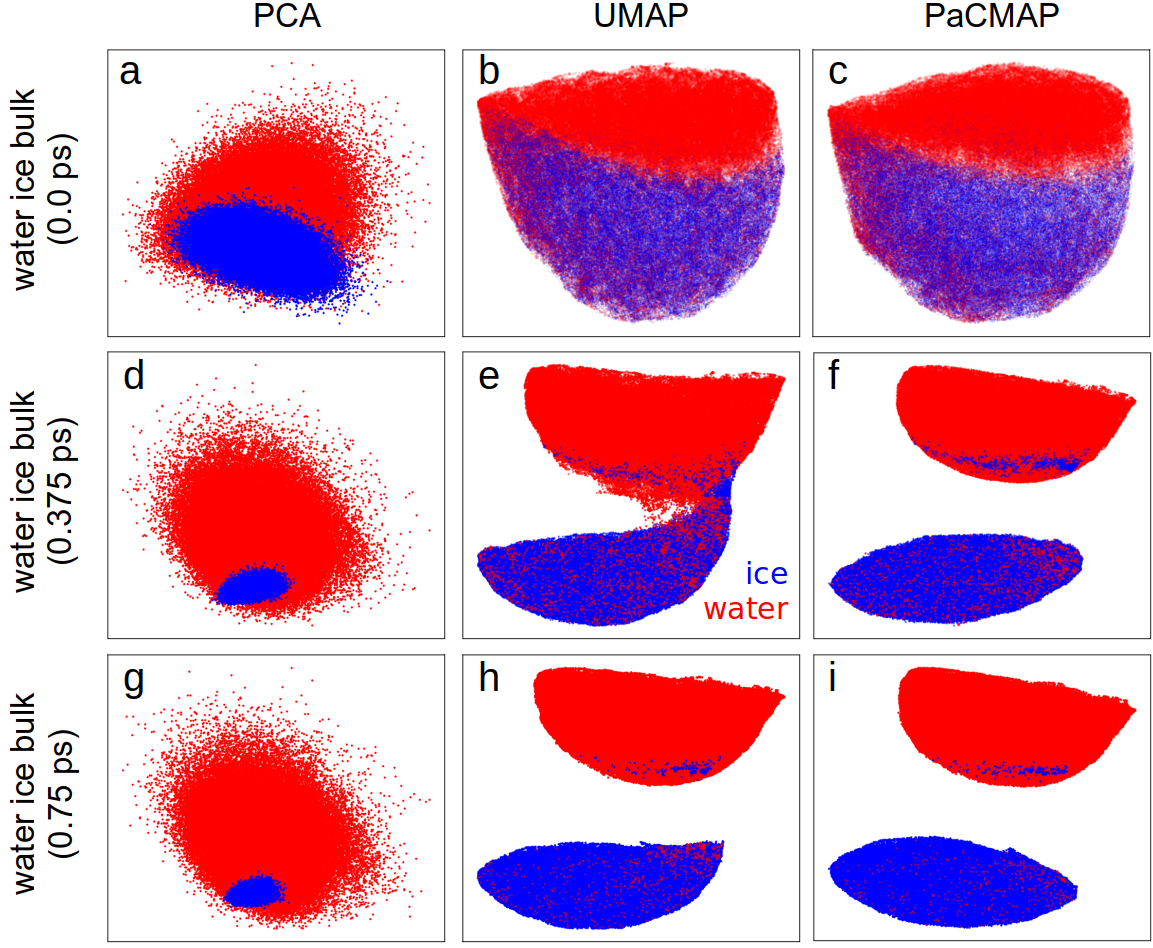}
\caption{
\textbf{Comparison of clustering results using PCA, UMAP, and PaCMAP algorithms for bulk water and ice.} Panels (a-c) show results without time averaging, (d-f) show results averaged over a 0.375 ps time interval, and (g-i) display results averaged over a 0.75 ps time interval.
}
\label{fig:water}
\end{figure*}

\begin{figure*}[ht]
\centering
\includegraphics[width=0.7\textwidth]{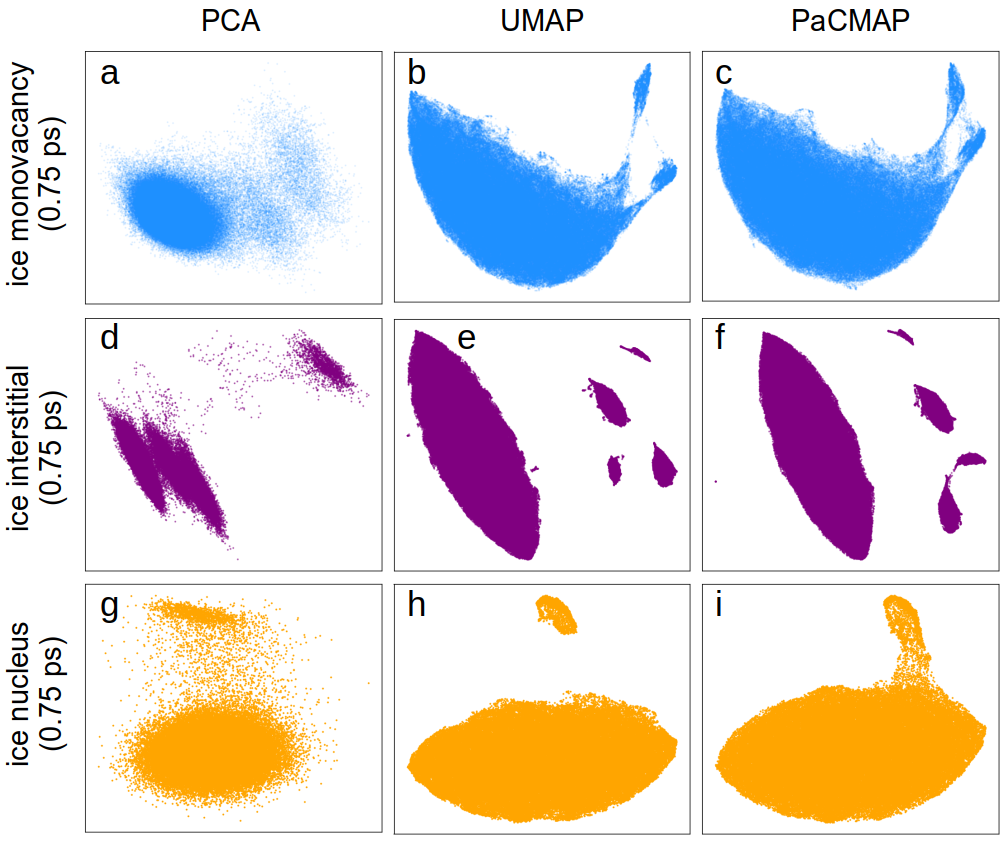}
\caption{
\textbf{Comparison of clustering results using PCA, UMAP, and PaCMAP algorithms for special local environments in ice.} Included cases are the ice monovacancy (a-c) and the ice interstitial (d-f) defects and the rounded ice nucleus in water (g-i), with potential descriptors averaged over a 0.75 ps time interval.
}
\label{fig:with}
\end{figure*}

The second part of the paper is focused on the differentiation of ice-like and liquid-like water molecules in the context of defects and ice nucleation. It would be expected that classifying disordered and ordered phases would be easier than between similar crystal polymorphs, but this task was more challenging than the previous case of silicon. Mid- to long-range ordering is a critical feature for differentiation and is not primarily addressed by our potential descriptors. As shown in Figs. \ref{fig:water}a-c, none of the tested methods distinguishes the water phase from the ice phase. This is due to the greater significance attributed to O-H bonds of water molecules compared to hydrogen bonds in used atomic potential descriptors. When the O-H bond length is fixed, as in the case of certain classical force fields, a complete separation between liquid and solid phases is possible. Thermal fluctuations of OH bonds can be reduced by applying a moving average to the molecular dynamics trajectory.

PaCMAP achieves complete water and ice phase separation at a moving average interval of 0.375 ps (Fig. \ref{fig:water}f), with classification errors of only 1.14\% for water and 0.68\% for ice. As the moving average length increases, phase differentiation accuracy improves. For instance, a 0.75 ps moving average reduces errors to 0.34\% for water and 0.14\% for ice. In contrast, UMAP fails to achieve complete cluster separation at 0.375 ps (Fig. \ref{fig:water}e) or even at 0.5 ps (Fig. S4). Only with a 0.75 ps moving average does UMAP entirely separate water and ice local environments (Fig. \ref{fig:water}h), albeit with a higher classification error for water (0.59\%) compared to PaCMAP (0.34\%). The classification error for ice remains similar, at 0.14\%.

The increased classification error for water is primarily due to the probability of water molecules forming local ice-like structures. In contrast, ice molecules remain closely aligned with their crystal lattice structure.

\subsubsection{Molecular defects}
Time averaging has been shown to be essential in order to differentiate between water and ice phases using UMAP. Specifically, a moving interval of at least 0.75 ps is required to separate water and ice phases for UMAP effectively. Therefore, we employed this same interval in subsequent tests to identify molecular monovacancy, interstitial atoms, and ice nuclei in the liquid phase.

For localizing monovacancies in ice, UMAP proved the most effective for separating clusters (Figs. \ref{fig:with}a-c), although complete separation is not achieved. The two emerging clusters correspond to distinct local chemical environments near monovacancy defects, specifically, two oxygen atoms missing a hydrogen donor and two oxygens missing a hydrogen acceptor per structure. These four oxygens near molecular monovacancy enable straightforward tracking of the monovacancy location.

In the case of identifying interstitial atoms, even PCA demonstrates sufficient clustering (Fig. \ref{fig:with}d). UMAP and PaCMAP not only identify interstitials in 99.3\% of structures (top right clusters in Figs. \ref{fig:with}e and \ref{fig:with}f) but also their neighbor atoms (three clusters below interstitial cluster). Notably, the small cluster on the left represents a second neighbor to an interstitial atom that has been displaced from its crystal position due to interstitial diffusion.

\subsubsection{Ice nucleus in supercooled water}
Our previous analysis compared the bulk structures of the water and ice phases. Identifying an ice nucleus within the liquid phase proved more challenging due to its distinct structure, influenced by interfacial stress. The interface is characterized by molecules with intermediate properties between liquid and solid, making it difficult to distinguish between them. To avoid finite-size effects in nucleation studies, the crystalline nucleus was significantly smaller than the surrounding liquid, resulting in an enormous class imbalance among molecules in the three phases (liquid, interface, and nucleus). Even using a moving average of 0.75 ps, only UMAP could effectively differentiate between the liquid and ice nucleus, as illustrated in Figs. \ref{fig:with}g-i. While PaCMAP showed a visible cluster forming in the upper right, it is not distinct. The structure contained approximately 75,000 molecules in the liquid phase and about 2,000 molecules in the solid phase, resulting in an enormous class imbalance. Augmenting the dataset with an additional 2,000 local environments of bulk ice structures enabled complete cluster separation even for PaCMAP (Fig. S5b). However, only partial overlap with the ice bulk environment is observed in the newly separated cluster. The non-overlapping segment corresponds to interface atoms, as indicated by the periodicity loss at the nucleus's edge in Fig. S5c.

\section{Discussion}
We have demonstrated the applicability of atomistic potential descriptors for unsupervised identification of local atomic environments in silicon and water systems. By utilizing these potential descriptors, we can not only classify phases but also identify point defects and interfaces. We compared the clustering capabilities of two widely used methods, PCA and UMAP, with a novel approach, PaCMAP. Although PCA provided limited valuable information for phase or defect detection, its initialization served as an effective starting point for both UMAP and PaCMAP. Without PCA initialization, the clustering results of both methods were unstable.

Both UMAP and PaCMAP successfully separated silicon phases and identified diamond point defects. UMAP outperformed PaCMAP in distinguishing the first neighbors of the silicon surface, while PaCMAP only identified the surface atoms. Furthermore, UMAP could identify ice nuclei in water without additional data points, whereas PaCMAP required augmenting the dataset. This sensitivity to class imbalance suggests that PaCMAP may struggle with unbalanced datasets. In contrast, when class imbalance was not an issue, such as distinguishing between water and hexagonal ice, PaCMAP outperformed UMAP. PaCMAP required only a 0.375 ps moving average, whereas UMAP needed twice that time.  Although hyperparameter tuning could potentially yield similar results for both methods, our focus was on unsupervised clustering, where the number of clusters is generally unknown a priori.

In conclusion, PaCMAP's superiority over UMAP has not been confirmed. While PaCMAP excelled in separating the water and ice phases, it struggled with unbalanced datasets. Although UMAP clustering is sensitive to hyperparameter variations, it performs effectively with default settings, except for the initialization process, where the default spectral initialization produces unstable results. Both algorithms perform well overall, and the choice between them should be determined by the specifics of the system under study.

\section{Methods}

\subsection{PaCMAP algorithm}
The overarching objective of PaCMAP is to bring proximate data points within the high-dimensional space into closer proximity within the low-dimensional space, as well as more distant data points in the original space to greater distances within the low-dimensional space. Since it would be computationally demanding to consider all the distances between all atoms with each other during optimization, PaCMAP restricts the number of neighbors for each data point to a finite value. Neighbors are categorized into three types: near pairs ($n_{NP}$), mid-near pairs ($n_{MN}$), and further pairs ($n_{FP}$). Attractive forces are applied to near and mid-near pairs, whereas repulsive forces are exerted on further pairs. The selection of neighbors is a one-time process, and the neighbor pairs remain constant throughout the optimization phase. The whole algorithm unfolds in three steps:

\subsubsection{Choosing neighbors}
To determine near neighbors, an initial step involves the selection of a subset comprising the minimum of either $n_{NP} + 50$ or the total number of observations ($N$) nearest neighbors based on Euclidean distance. Subsequently, the nearest neighbors are identified according to the scaled distance metric $d_{ij}^2 = \frac{||\textbf{x}_i-\textbf{x}_j||^2} {\sigma_i \sigma_j}$ between observation pairs (i, j), where $\textbf{X}_i$ represents atomistic potential descriptors and $\sigma_i$ is the average distance between observation $i$ and its Euclidean nearest neighbors falling in the fourth to sixth positions. This scaling is implemented to accommodate potential variations in the magnitudes of neighborhoods across different regions of the feature space. In selecting mid-near pairs, six additional points are randomly sampled (uniformly), and the second nearest point among them is chosen. Lastly, further pairs are determined by randomly selecting $n_{FP}$ additional points.

\subsubsection{Initialize low dimension embedding $\textbf{Y}$} 
 Each datapoint $i$ is associated with a 2D vector $\textbf{y}_i$. As PaCMAP is a nonparametric algorithm without an underlying function that generates low-dimension embedding, values are generated by selecting the most important PCA components.


\subsubsection{Minimilize loss}
With a predefined neighbor list for each datapoint and an initial position in the low-dimensional embedding provided by PCA, the optimization process commences minimizing the loss function by varying $\textbf{y}_i$ with the ADAM optimizer \cite{adam}:
\begin{equation}
\small
L =\sum_i^N \left( w_{NP} \sum_j^{n_{NP}}\frac{\tilde{d}_{ij}}{10+\tilde{d}_{ij}}+w_{MN}\sum_k^{n_{MN}}\frac{\tilde{d}_{ik}}{10000+\tilde{d}_{ik}} + w_{FP}\sum_l^{n_{FR}} \frac{1}{1+\tilde{d}_{il}}\right)
\end{equation}
where $\tilde{d}_{ab} = ||\textbf{y}_a - \textbf{y}_b||^2 + 1$. The first two terms correspond to the attractive interactions between near and mid-near pairs. The third term represents a repulsive interaction that facilitates cluster separation. The pairs contribute to the loss function with weights determined by the coefficients $w_{NP}$, $w_{MN}$, and $w_{FP}$, collectively constituting the overall loss. These weights are dynamically updated throughout the algorithm as part of the optimization process. The dynamic update of weights follows a specific scheme during different iterations of the optimization process:

\begin{itemize}
\item  Iterations 0 to 100: $w_{NP}=2$, \\$w_{MN} = 1000\left(1-\frac{t-1}{100}\right) + 3\left(\frac{t-1}{100}\right)$, $w_{FP} = 1$ 
\item  Iterations 101 to 200: $w_{NP}=3$, $w_{MN} = 3$, $w_{FP} = 1$ 
\item Iterations 201 to 450: $w_{NP}=1$, $w_{MN} = 0$, $w_{FP} = 1$ 
\end{itemize}

The optimization process has three distinct phases designed to circumvent local optima. The initial phase focuses on global structure achieved through substantial weighting of mid-near pairs. As this phase progresses, weights on mid-near pairs gradually decrease, facilitating algorithmic focus shifting from global to local structures. The subsequent phase concentrates on enhancing local structure while retaining the global structure acquired in the first phase. Finally, the third phase prioritizes the refinement of local structure by reducing the weight of mid-near pairs to zero, accentuating the role of repulsive forces to separate cluster boundaries more distinctly. Fig. S6 provides an example of the optimization process for silicon phases. 

\subsection{Parametric PaCMAP}
In this section, we address the most significant limitation of the PaCMAP algorithm: its nonparametric nature. It means that predicting new data points is not directly possible, as the algorithm does not construct a function $f(x):R^d \rightarrow R^2$  that would map high-dimensional descriptors to a 2D latent space. This limitation can be a considerable constraint in certain real-world applications where the goal is to cluster training data and apply the learned mapping to new, simulated data points. A practical approach involves generating embedding/labels, which are then employed as output for supervised regression or classification tasks. Subsequently, a simple feedforward neural network can be easily trained and rapidly produce predictions for new local environments. To illustrate this approach, we trained a neural network with three hidden layers, each comprising 50 nodes and employing ReLU activation functions for silicon and bulk liquid and ice phases.

\begin{figure}[ht]
\centering
\includegraphics[width=6cm]{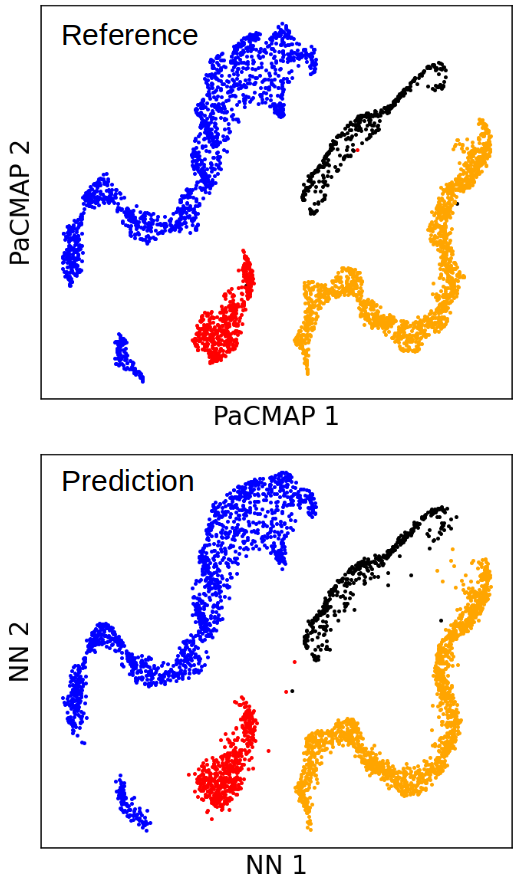}
\caption{
\textbf{Parametric prediction of PaCMAP embedding.} Comparison of reference test data (top) and neural network predictions (bottom) for distinguishing between silicon phases. The test dataset contains 4,000 local chemical environments.
}
\label{fig:nn}
\end{figure}

In Fig. \ref{fig:nn}, we compare the neural network's predictions with the test reference dataset (PaCMAP mapping), proving visually that the neural network has successfully learned the underlying mapping of PaCMAP. A similar comparison of bulk liquid and ice phases is provided in Fig. S7. However, classification is a more practical approach for real-world applications. Each cluster from the training data would be assigned a class, and a classifier would then predict the class of each cluster directly from the descriptors, eliminating the need for 2D embedding.


\subsection{Datasets, atomistic potential descriptors, and PaCMAP hyperparameters}
We have selected two well-known systems for benchmarking unsupervised algorithms: silicon (Si) and water (H$_2$O). In the case of silicon, our focus is on identifying the crystal lattice, point defects, and the surface. Our emphasis for water molecules is on liquid and hexagonal ice distinction and molecular point defects.

The General-Purpose Interatomic Potential for Silicon encompasses 2,475 structures, which can be categorized into 23 different structure types \cite{Bartk2018}. Not all structure types contain a sufficient number of structures for data-driven analysis. Therefore, we restrict our analysis to distinguishing between diamond, $\beta$-Sn, simple hexagon (SH), and liquid phase, but also localization of monovacancies, interstitial positions, and identifying surface atoms.

 Water structures include bulk liquid, ice, and a spherical nucleus surrounded by supercooled water. All structures have been generated by running MD simulations with n2p2 \cite{Singraber2019} -- LAMMPS \cite{Thompson2022} using a Behler-Parrinello neural network trained on ab initio data based on the RPBE-D3 zero damping density functional \cite{Morawietz2016}. 
 Initial ice-Ih structures were generated with GenIce \cite{matsumoto2018genice} and then simulated in the anisotropic NpT ensemble at 250 K and 0~bar for 700 ps. Supercooled water presents prolonged relaxation, so producing equilibrated structures requires either very long simulations or the use of special techniques. Here, we have run a 32 replica parallel tempering simulation at constant 0 bar pressure during 8 ns covering from 211 K up to 335 K. Then, the last structure from the distribution of 250~K was selected and simulated for 2~ns more at constant temperature and pressure (250 K and 0 bar) for production-level data acquisition. To produce the nucleus system, we inserted a perfect spherical ice-Ih nucleus of about 2,000 water molecules into a supercooled water configuration, producing 78,856 water molecules. We pre-equilibrated the system using the TIP4P/Ice force field \cite{abascal2005potential} via the GROMACS package \cite{abraham2015gromacs}. The temperature was set to 250 K, pressure to 0 bar, and the nucleus size barely changed during 500~ps. Then, we switched to the MLP and slightly heated the system towards 260 K for 10~ps. Finally, we have run 2.5~ns in the NpH ensemble for production-level structures at 0 bar. The nucleus remained relatively stable in size, and the average temperature quickly converged to $\sim$ 255 K. 

 We employ Behler-Parinello descriptors (symmetry functions) to encode the local structure information due to their simplicity and widespread applicability \cite{Behler2007}. 
 For silicon, we constructed a descriptor vector comprising 20 radial Gaussian symmetry functions and 15 angular polynomial symmetry functions \cite{Bircher2021} (details are provided in Tables S1 and S2). These symmetry functions are designed generally without leveraging specific knowledge about the bonding \cite{Gastegger2018}. A cutoff distance of 5 Å was set to prevent self-interaction in smaller supercells within the dataset. For water analysis, we chose well-tested symmetry functions commonly employed in molecular dynamics simulations accelerated by MLPs \cite{Morawietz2016} with a cutoff of 6.35 \AA. Only oxygen descriptors that include both oxygen-oxygen and oxygen-hydrogen interactions have been utilized in analyses. 

Each data point corresponds to an individual atomic environment within the supercell without averaging over the cell or time unless otherwise specified.

Regarding the hyperparameters of the PaCMAP algorithm, we have fixed parameters for all presented cases. This entails $n_{NP} = 10, n_{MN} = 5,$ and $ n_{FP} = 25$ with PCA initialization.

\subsection{Software}
 Production-level molecular dynamics simulations are performed with n2p2 \cite{Singraber2019} -- LAMMPS \cite{Thompson2022}. GROMACS \cite{abraham2015gromacs} is used for setting up the ice-Ih nucleus system. Initial configurations of ice are generated with GenIce \cite{matsumoto2018genice}. The symmetry functions are evaluated using the n2p2 package \cite{Singraber2019}. The structures are rendered using OVITO \cite{ovito}. Dimensionality reduction is performed using packages: scikit-learn for PCA \cite{scikit-learn}, UMAP \cite{McInnes2018}, and PaCMAP \cite{PaCMAP}. Neural networks trained on PaCMAP labels are constructed and interfered with PyTorch \cite{pytorch}. 

\section*{Data availability}
The atomic descriptors and corresponding structures are publicly available on Zenodo \cite{zenodopacmap}.

\section*{Acknowledgments}
We acknowledge financial support by the Doctoral College Advanced Functional Materials – Hierarchical Design of Hybrid Systems DOC 85 doc.funds and SFB-TACO 10.55776/F81 funded by the Austrian Science Fund (FWF) and by the Vienna Doctoral School in Physics (VDSP). The computational results presented have been partly achieved using the Vienna Scientific Cluster (VSC).

\section*{Competing interests}
The authors declare no competing interests

\section*{Contributions}

LK implemented the methodology and conducted the clustering analyses.
PMdH prepared the water datasets. 
CD supervised and provided funding acquisition and resources.  All authors participated in the conceptualization of the project and the validation of the results. All authors participated in preparing, writing, and editing the manuscript.

\section*{Additional information}

\textbf{Supplementary information} includes complementary analyses and a list of the atomistic potential descriptors for silicon. 

\textbf{Correspondence} and requests for details should be addressed to Christoph Dellago.

\section*{References}
\bibliography{bibtex}

\end{document}


\section*{Supporting Information for Publication: Unsupervised identification of crystal defects from atomistic potential descriptors}
\centering
Lukáš Kývala, Pablo Montero de Hijes, Christoph Dellago\\
\textit{University of Vienna, 1090 Vienna, Austria}
\renewcommand{\arraystretch}{1.1}

\vspace{2cm}

\begin{figure}[ht]
\centering
\includegraphics[width=0.7\textwidth]{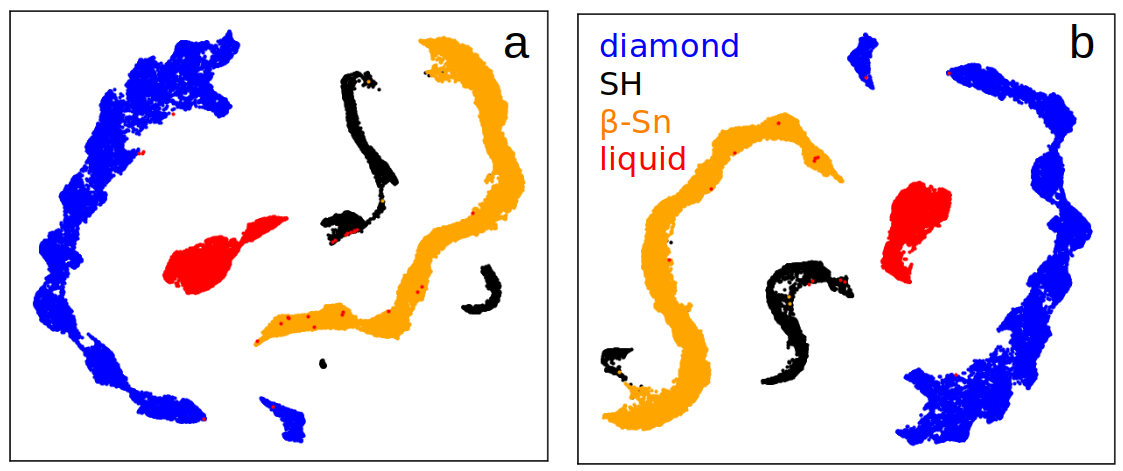}
\caption{UMAP clustering for silicon phases with default spectral initialization. a) The SH phase (black) is incorrectly divided into three clusters. b) A portion of the SH phase (black) is merged with the $\beta$-Sn cluster (orange). Note that two blue clusters, also observed in PaCMAP, correctly represent the diamond phase but correspond to significantly different densities.}
\label{fig:umap}
\end{figure}

\begin{figure*}[ht]
\centering
\includegraphics[width=1.0\textwidth]{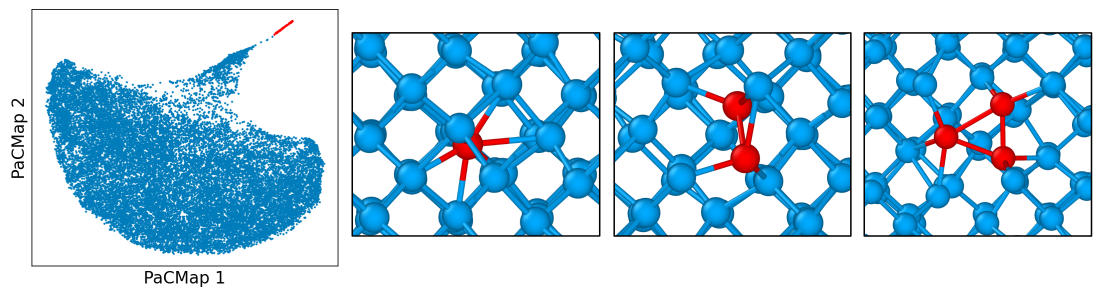}
\caption{PaCMAP clustering of interstitials in the diamond silicon structure, showing three randomly chosen structures with atoms colored by their corresponding clusters.}
\label{fig:inter}
\end{figure*}

\begin{figure}[ht]
\centering
\includegraphics[width=0.7\textwidth]{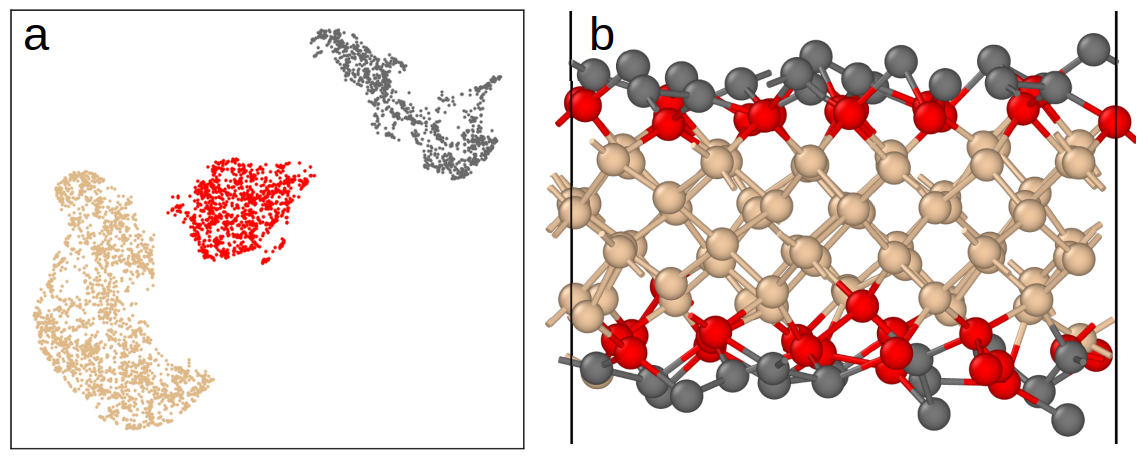}
\caption{(a) UMAP clustering of silicon's diamond (001) surface. (b) Example of a silicon structure with atoms colored according to their UMAP clusters.}
\label{fig:surface}
\end{figure}

\begin{figure}[ht]
\centering
\includegraphics[width=1.0\textwidth]{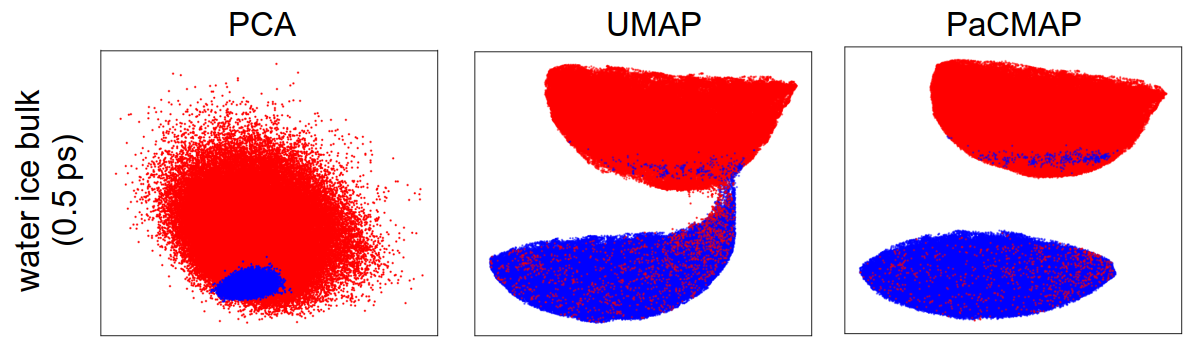}
\caption{Comparison of clustering results using PCA, UMAP, and PaCMAP algorithms for water and ice bulk phases, with potential descriptors averaged over a 0.50 ps time interval.}
\label{fig:water050}
\end{figure}

\begin{figure}[ht]
\centering
\includegraphics[width=1.0\textwidth]{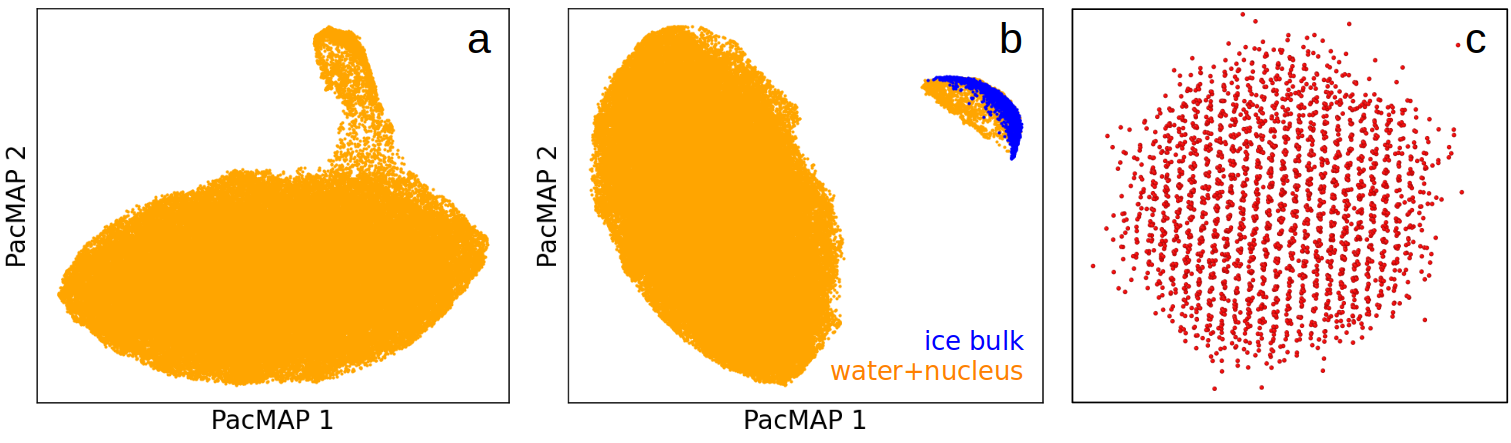}
\caption{PaCMAP clustering of a water supercell: (a) with an ice nucleus, (b) with 2,000 additional ice bulk environments, and (c) highlighting the identified ice nucleus. Potential descriptors are averaged over a 0.75 ps time interval.}
\label{fig:nucleus}
\end{figure}

\begin{figure}[ht]
\centering
\includegraphics[width=1.0\textwidth]{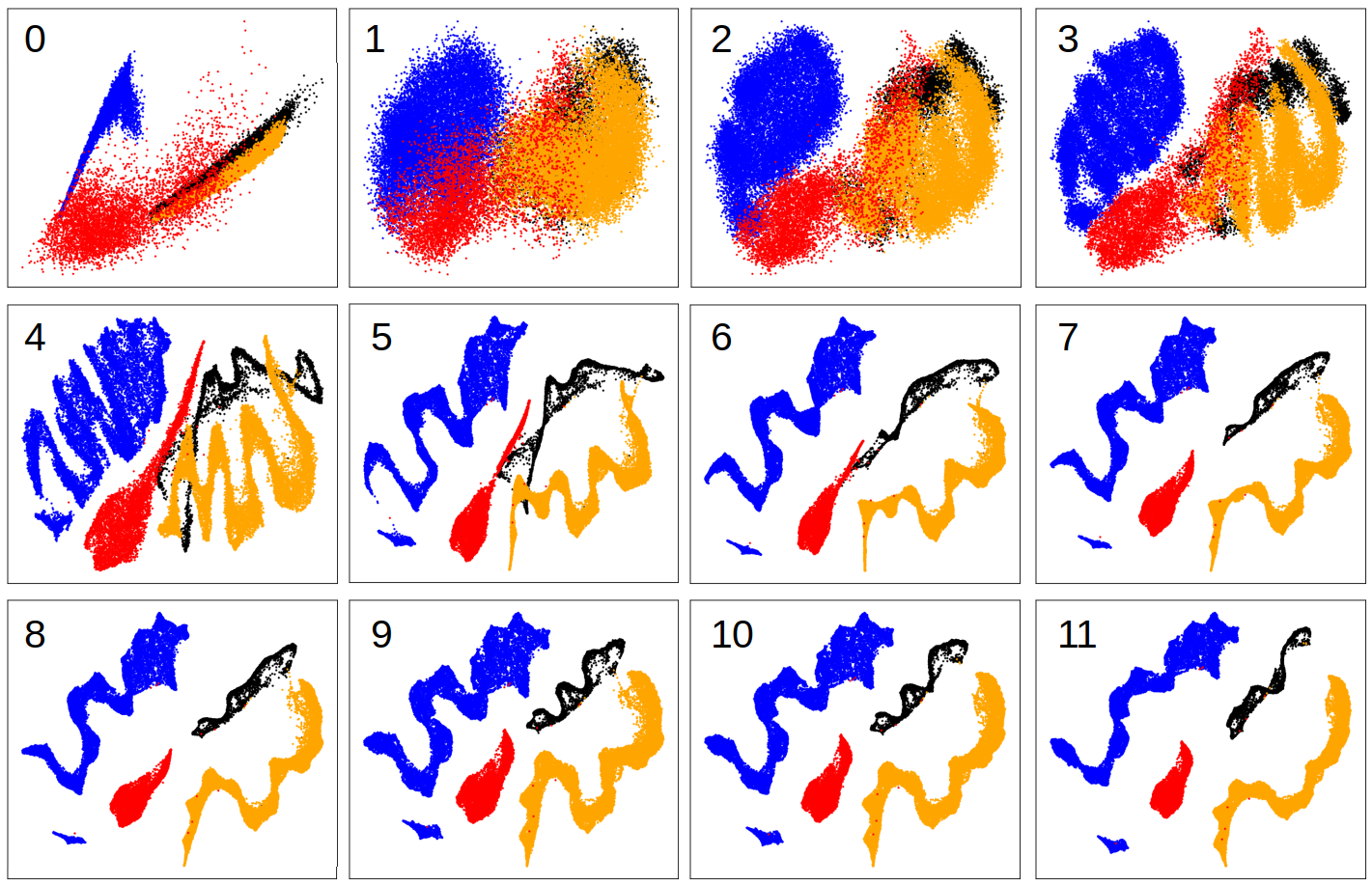}
\caption{Visualizations of the low-dimensional embedding during the optimization process of PaCMAP on the silicon dataset. Iteration 0 shows the PCA initialization.}
\label{fig:optimalization}
\end{figure}

\begin{table}[ht]
\footnotesize
\caption{Parameters of the radial symmetry functions for silicon.}
\centering
\begin{tabular}{lccc}
G$^{rad}$   & r$_{min}$ (Å)& r$_{max}$ (Å) & $\theta_{min}$ (\textdegree) \\
     \hline \hline
&	0.0556	&	0.0	&	5.0	\\
&	0.0499	&	0.0	&	5.0	\\
&	0.0450	&	0.0	&	5.0	\\
&	0.0408	&	0.0	&	5.0	\\
&	0.0372	&	0.0	&	5.0	\\
&	0.0340	&	0.0	&	5.0	\\
&	0.0313	&	0.0	&	5.0	\\
&	0.0288	&	0.0	&	5.0	\\
&	0.0266	&	0.0	&	5.0	\\
&	0.0247	&	0.0	&	5.0	\\
&	18.0	&	3.000	&	5.0	\\
&	18.0	&	3.167	&	5.0	\\
&	18.0	&	3.333	&	5.0	\\
&	18.0	&	3.500	&	5.0	\\
&	18.0	&	3.667	&	5.0	\\
&	18.0	&	3.833	&	5.0	\\
&	18.0	&	4.000	&	5.0	\\
&	18.0	&	4.167	&	5.0	\\
&	18.0	&	4.333	&	5.0	\\
&	18.0	&	4.500	&	5.0	\\

     \hline
\end{tabular}
\end{table}

\begin{table}[ht]
\footnotesize
\caption{Parameters of the angular symmetry functions for silicon.}
\centering
\begin{tabular}{lcccc}
G$^{ang}$  & r$_{min}$ (Å)& r$_{max}$ (Å) & $\theta_{min}$ (\textdegree) & $\theta_{max}$ (\textdegree)\\
     \hline \hline
    & -3   & 3    & -180   & 180    \\
    & -3   & 3    & 0   & 360    \\
    & -3   & 3    & 0   & 180    \\
    & -3   & 3    & 0   & 120    \\
    & -3   & 3    & 60   & 180    \\
    & -4   & 4    & -180   & 180    \\
    & -4   & 4    & 0   & 360    \\
    & -4   & 4    & 0   & 180    \\
    & -4   & 4    & 0   & 120    \\
    & -4   & 4    & 60   & 180    \\
    & -5   & 5    & -180   & 180    \\
    & -5   & 5    & 0   & 360    \\
    & -5   & 5    & 0   & 180    \\
    & -5   & 5    & 0   & 120    \\
    & -5   & 5    & 60   & 180    \\

\end{tabular}
\end{table}

\begin{figure}[ht]
\centering
\includegraphics[width=12cm]{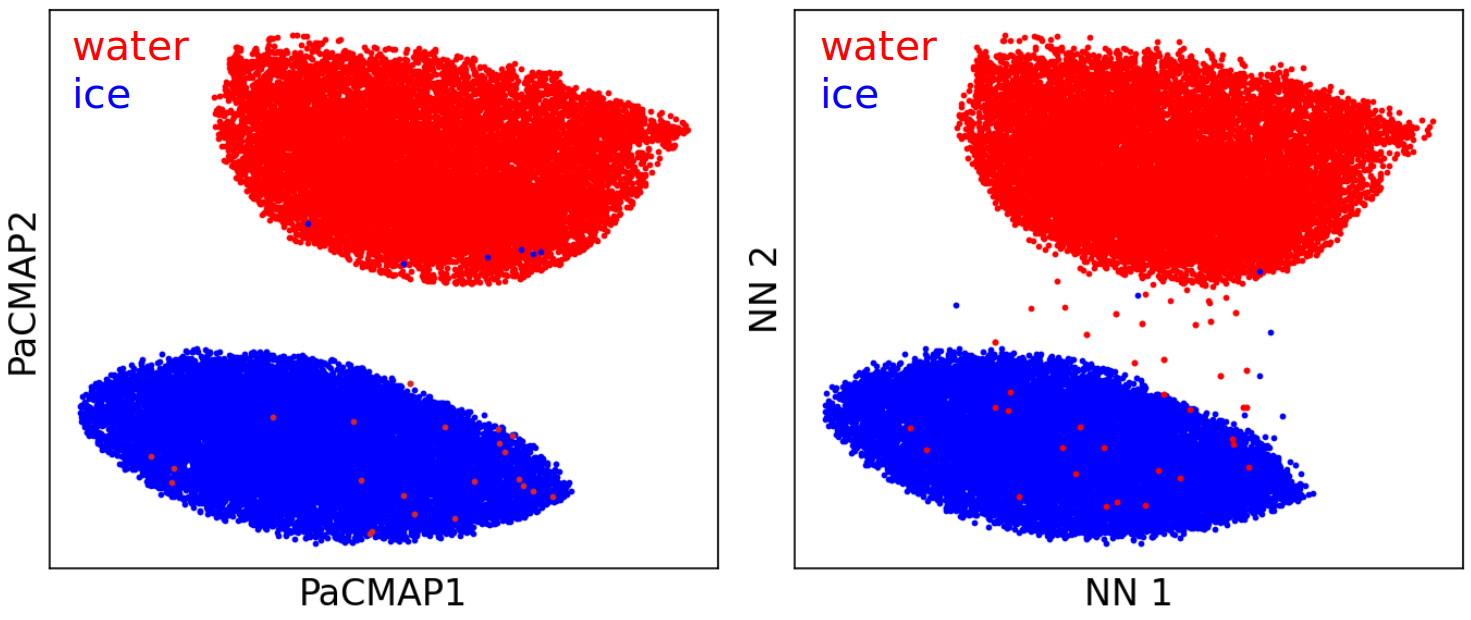}
\caption{Comparison of reference test data (left) and neural network predictions (right) for distinguishing between water and ice. The test dataset comprises 30,000 molecules.}
\label{fig:nn}
\end{figure}